\documentclass[twocolumn]{aastex62}
\usepackage{amsmath}
\graphicspath{{./}{figures/}}

\shorttitle{Localizing The First Interstellar Meteor With Seismometer Data}
\shortauthors{Siraj \& Loeb}

\begin{document}

\title{\large Localizing The First Interstellar Meteor With Seismometer Data}

\email{amir.siraj@cfa.harvard.edu, aloeb@cfa.harvard.edu}

\author{Amir Siraj}
\affil{Department of Astronomy, Harvard University, 60 Garden Street, Cambridge, MA 02138, USA}

\author{Abraham Loeb}
\affiliation{Department of Astronomy, Harvard University, 60 Garden Street, Cambridge, MA 02138, USA}




\begin{abstract}
The first meter-scale interstellar meteor (IM1) was detected by US government sensors in 2014, identified as an interstellar object candidate in 2019, and confirmed by the Department of Defense (DoD) in 2022. We use data from a nearby seismometer to localize the fireball to a $\sim 16 \mathrm{\; km^2}$ region within the $\sim 120 \mathrm{\; km^2}$ zone allowed by the precision of the DoD-provided coordinates. The improved localization is of great importance for a forthcoming expedition to retrieve the meteor fragments.

\end{abstract}



\section{Introduction}

The first interstellar meteor (IM1), CNEOS\footnote{\url{https://cneos.jpl.nasa.gov/}} 2014-01-08, was detected by U.S. Department of Defense (DoD) sensors through the light that it emitted as it burned up in the Earth's atmosphere off of the coast of Papua New Guinea in 2014 \citep{2022ApJ...939...53S}. The material strength of IM1 appears to be higher than all other 272 meteors in the CNEOS catalog \citep{2022RNAAS...6...81S}. A forthcoming expedition aims to recover fragments from the ocean floor with an expedition to Papua New Guinea \citep{2022arXiv220800092S, 2022arXiv221200839T}. The area associated with the DoD-reported localization box for the fireball is $\sim 120 \mathrm{\; km^2}$, but given practical constraints the expedition cannot search an area that large. Hence, it is crucial to improve the precision of the fireball localization. In this \textit{Letter}, we use data from a nearby seismometer, AU MANU, to set new constraints on the fireball location.

\section{Method of Calculation}

We use the National Weather Service's Global Ensemble Forecast System model atmospheric profile (pressure as a function of temperature) for the DoD-reported fireball location, $(-1.3^{\circ}, 147.6^{\circ})$ at the time 2014-01-08 17:05:34 (private communication, Tim Gallaudet \& Steve Levine, April 27, 2022). We compute the sequential altitude differences as $dz = H(dP/P)$, where $H = (RT/g)$ is the scale height, $R$ is the gas constant, $g$ is the gravitational acceleration, $T$ is the temperature as a function of pressure $P$, and $dP$ are the sequential pressure differences. The sound speed as a function of temperature is given by, $v_a = 331.3 \mathrm{\; m \; s^{-1}} \; [1 + (T/546.3 \mathrm{\; C})]$. For this particular atmospheric profile, the sound speed as a function of altitude $z$ can be expressed as, $v_a = (347.2 - \frac{1.861 z}{\mathrm{km}}) \mathrm{\; m \; s^{-1}}$.

\begin{figure}
 \centering
\includegraphics[width=1\linewidth]{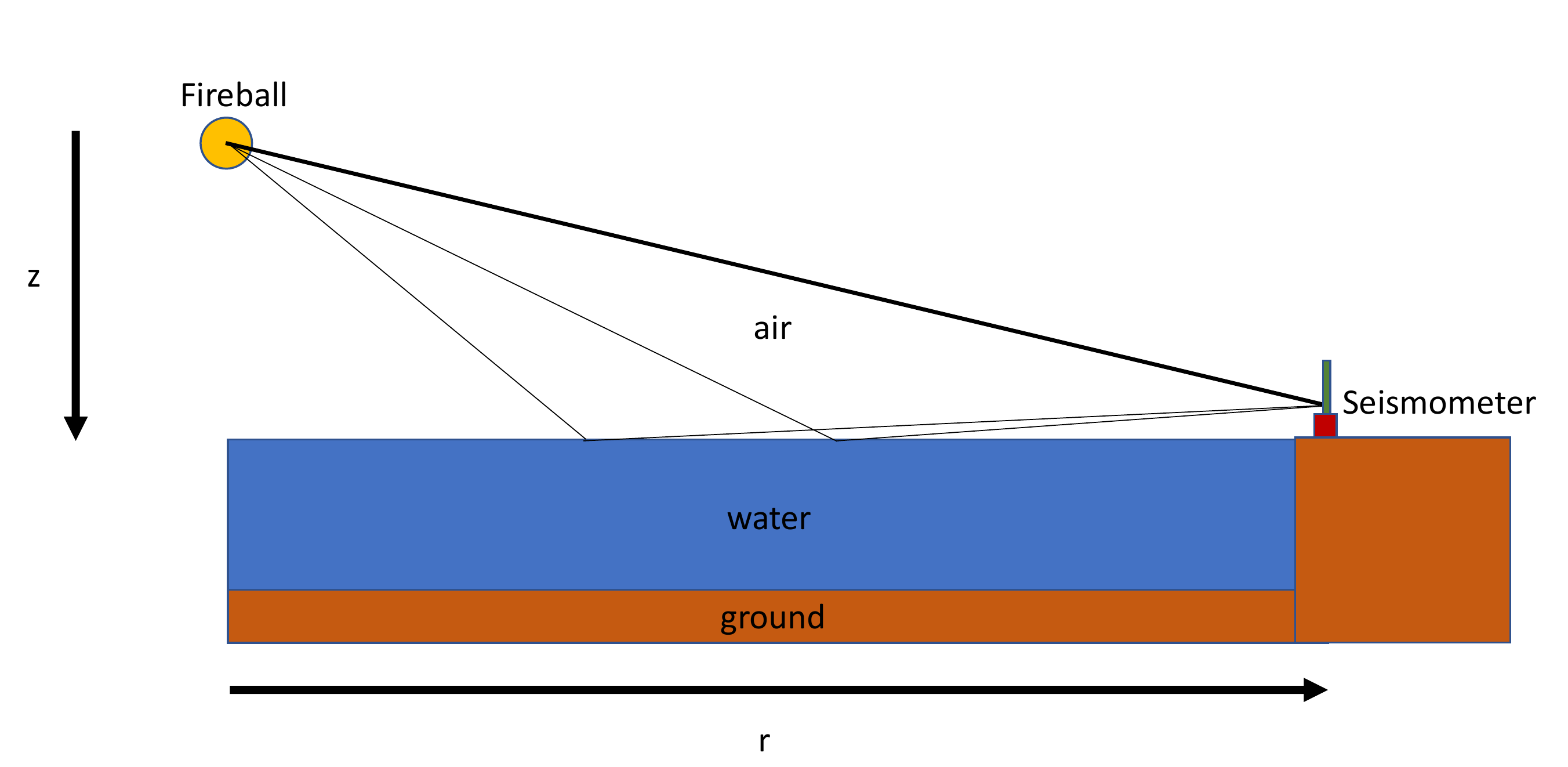}
\caption{Geometry of the different paths of sound waves from the meteor fireball to the seismometer: travelling directly through the air and reflecting off of the surface of the ocean. By summing over the different paths, we are able to fit the recorded sound signal for particular values of the horizontal distance ($r$) and altitude above sea level ($z$).}
\label{fig:diagram}
\end{figure}

\begin{figure}
 \centering
\includegraphics[width=1\linewidth]{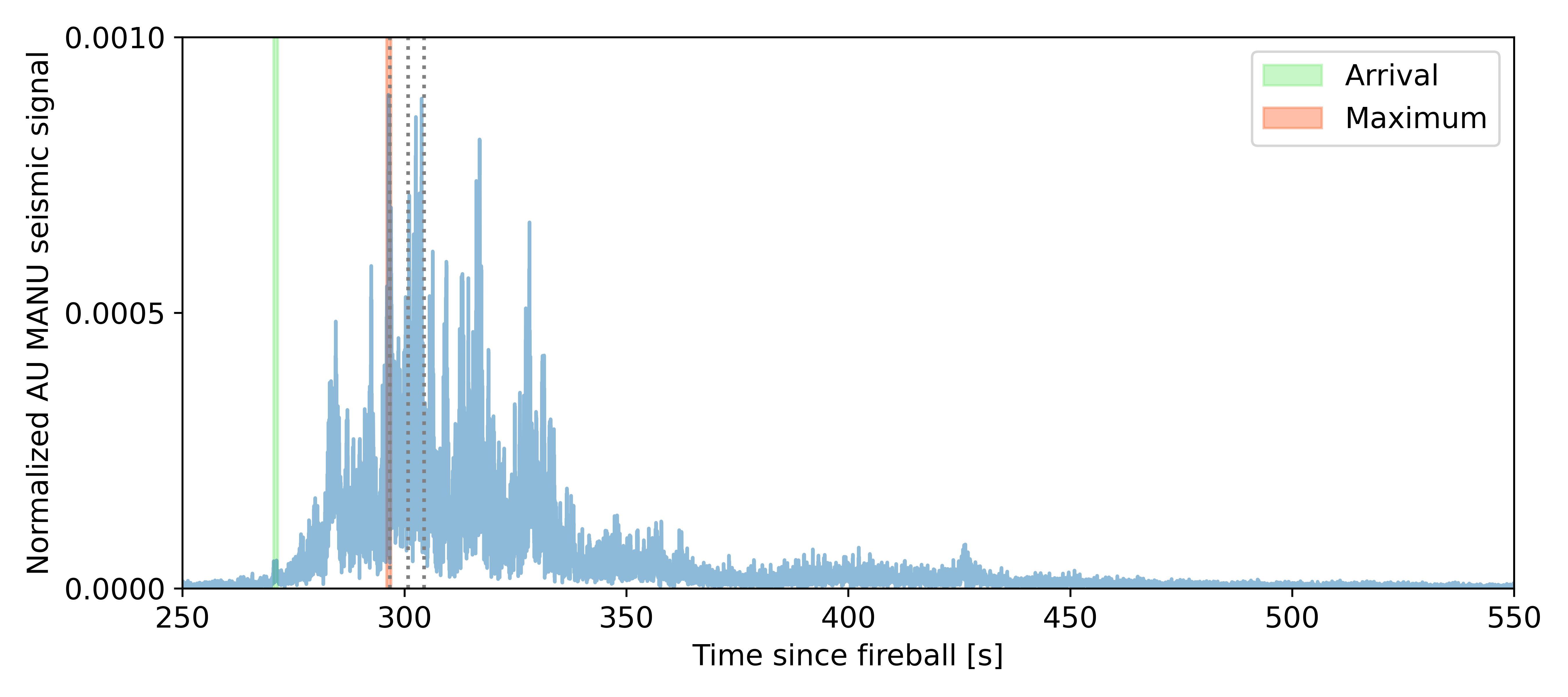}
\caption{AU MANU seismic signal as a probability function normalized to a unit area. The green region ($270.5 - 271.5 \mathrm{\;s}$ after the fireball) indicates the constraint for the arrival of the air-mediated sound waves, and the orange region ($296 - 297 \mathrm{\;s}$ after the fireball) illustrates the constraint for the peak signal produced by the air-mediated sound waves. The first, second, and third gray lines correspond to the third, second, and third peaks in the IM1 light curve \citep{2022RNAAS...6...81S}.}
\label{fig:seismic}
\end{figure}

The geometric setup is shown in Figure \ref{fig:diagram} and the AU MANU seismic signal that appears to correspond to the fireball (with the three directions of the seismometer added in quadrature, and the overall signal normalized to unity) is displayed in Figure \ref{fig:seismic}. We conservatively regard a packet of seismic energy that appears a few minutes before the one shown in Figure \ref{fig:seismic} as background noise because there is no clear mechanism by which a signal from the fireball would arrive at the AU MANU sensor so quickly, and given the prevalence of apparent seismic activity in the hours before and after the fireball, it is consistent with background noise.

For the isolated seismic signal from AU MANU, we require the first arrival of air-mediated sound waves from the fireball to occur at a delay of $270.5 - 271.5 \mathrm{\; s}$, the green region in Figure \ref{fig:seismic} corresponding to where the signal rises above the noise threshold. This implies sound waves traversing the shortest path in between the blast and the seismometer, a distance of $\sqrt{r^2 + z^2}$, where $z$ is the altitude of the fireball above the ocean surface and $r$ is the horizontal separation of the fireball from the seismometer. Similarly, we require the peak amplitude of the air-mediated sound waves to occur at a delay of $296 - 297 \mathrm{\; s}$ (within the orange region), corresponding to the apparent peak in the signal. To first order, this should correspond to sound waves that traveled a distance of roughly $r + (z / \sqrt{3})$ in the air (note that we use this approximation for the analytical derivation, but then verify the fit numerically).  

We solve for the following constraints,
\begin{align}
  \mathrm{Sound \; speed:} \; & v = (347.2 - \frac{1.861 z}{\mathrm{km}}) \mathrm{\; m \; s^{-1}} \label{eq1} \\
  \mathrm{Signal \; arrival:} \; & 270.5 \mathrm{\; s} < \frac{\sqrt{r^2 + z^2}}{v_a} < 271.5 \mathrm{\; s} \\ 
  \mathrm{Signal \; maximum:} \; & 296 \mathrm{\; s} < \frac{r + z/\sqrt{3}}{v_a} < 297 \mathrm{\; s}, \label{eq3} 
\end{align}
and find that this yields a ground distance $r = 83.9 \pm 0.7 \mathrm{\; km}$ and an altitude of $z = 16.9 \pm 0.8 \mathrm{\; km}$. Both values are consistent with the DoD-provided measurements of the fireball's 3D location.

There were three peaks observed in the lightcurve of IM1, with separations of $0.101 \mathrm{\; s}$ and $0.112 \mathrm{\; s}$ in time, and $2.0 \mathrm{\; km}$ and $2.3 \mathrm{\; km}$ in altitude, between the first and second peaks, and second and third peaks, respectively. The meteor's direction of motion was perpendicular to the vector extending to AU MANU, so the x-y motion does not affect the timing of acoustic transmission from the fireball to the seismometer. The signal maximum between $296 \mathrm{\; s}$ and $297 \mathrm{\; s}$ presumably corresponds to the third peak, because the signal from the third peak should have arrived earliest to AU MANU, given its relatively low altitude. It is also the strongest spike in the seismic signal; similarly, the third peak in the IM1 lightcurve was the most energetic. The altitude and temporal differences between the lightcurve peaks imply temporal separations for the acoustic maxima of $\sim [(2 \mathrm{\; km})/(\sqrt{3} \times 315 \mathrm{\;m /s})] - 0.101 \mathrm{\; s} = 3.57 \mathrm{\; s}$ and $\sim [(2.3 \mathrm{\; km})/(\sqrt{3} \times 315 \mathrm{\;m /s})] - 0.112 \mathrm{\; s} = 4.11 \mathrm{\; s}$ between the first and second peaks, and second and third peaks, respectively. Overall, this suggests that the maximum acoustic signatures associated with the second and first lightcurve peaks should arrive $4.11 \mathrm{\; s}$ and $7.68 \mathrm{\; s}$, respectively, after the signature associated with the third lightcurve peak. We illustrate this in Figure \ref{fig:seismic} as gray lines overlaid onto the AU MANU seismic signal, which shows good agreement. Our model implies that the third peak in the lightcurve corresponds to an altitude of $z = 16.9 \pm 0.9 \mathrm{\; km}$, the second to an altitude of $z = 19.2 \pm 0.9 \mathrm{\; km}$, and the first to an altitude of $z = 21.2 \pm 0.9 \mathrm{\; km}$. The resulting ram pressures are $244 \pm 27 \mathrm{\; MPa}$, $183 \pm 21 \mathrm{\; MPa}$, and $143 \pm 16 \mathrm{\; MPa}$, respectively.

\begin{figure}
 \centering
\includegraphics[width=1\linewidth]{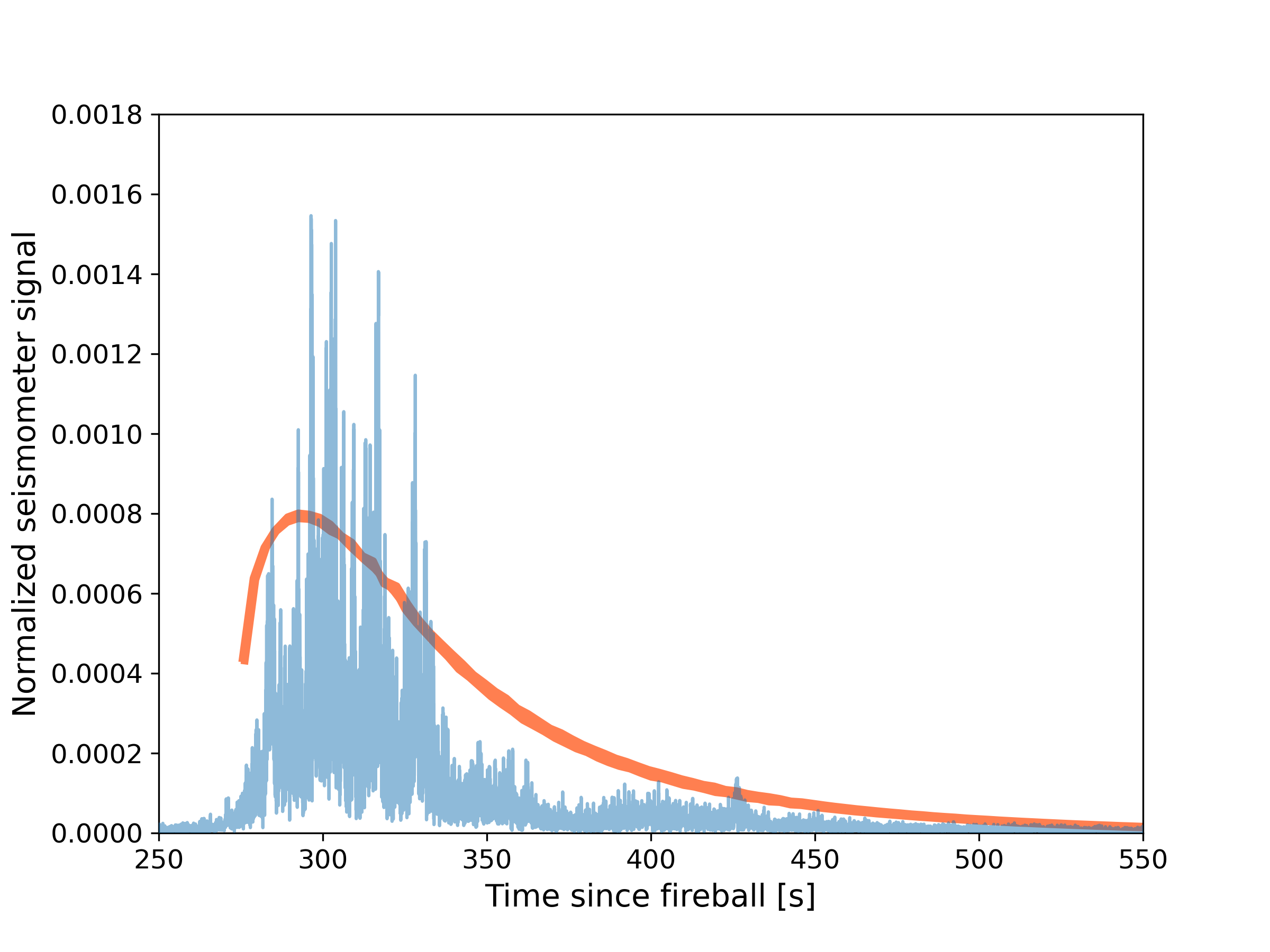}
\caption{The second packet of the seismic signal, normalized to unity. The orange curve indicates the result of the sound wave reflection model described in the text for a ground distance of $r = 83.9 \pm 0.7 \mathrm{\; km}$ and an altitude of $z = 16.9 \pm 0.9 \mathrm{\; km}$.}
\label{fig:peak2}
\end{figure}

\begin{figure}
 \centering
\includegraphics[width=1\linewidth]{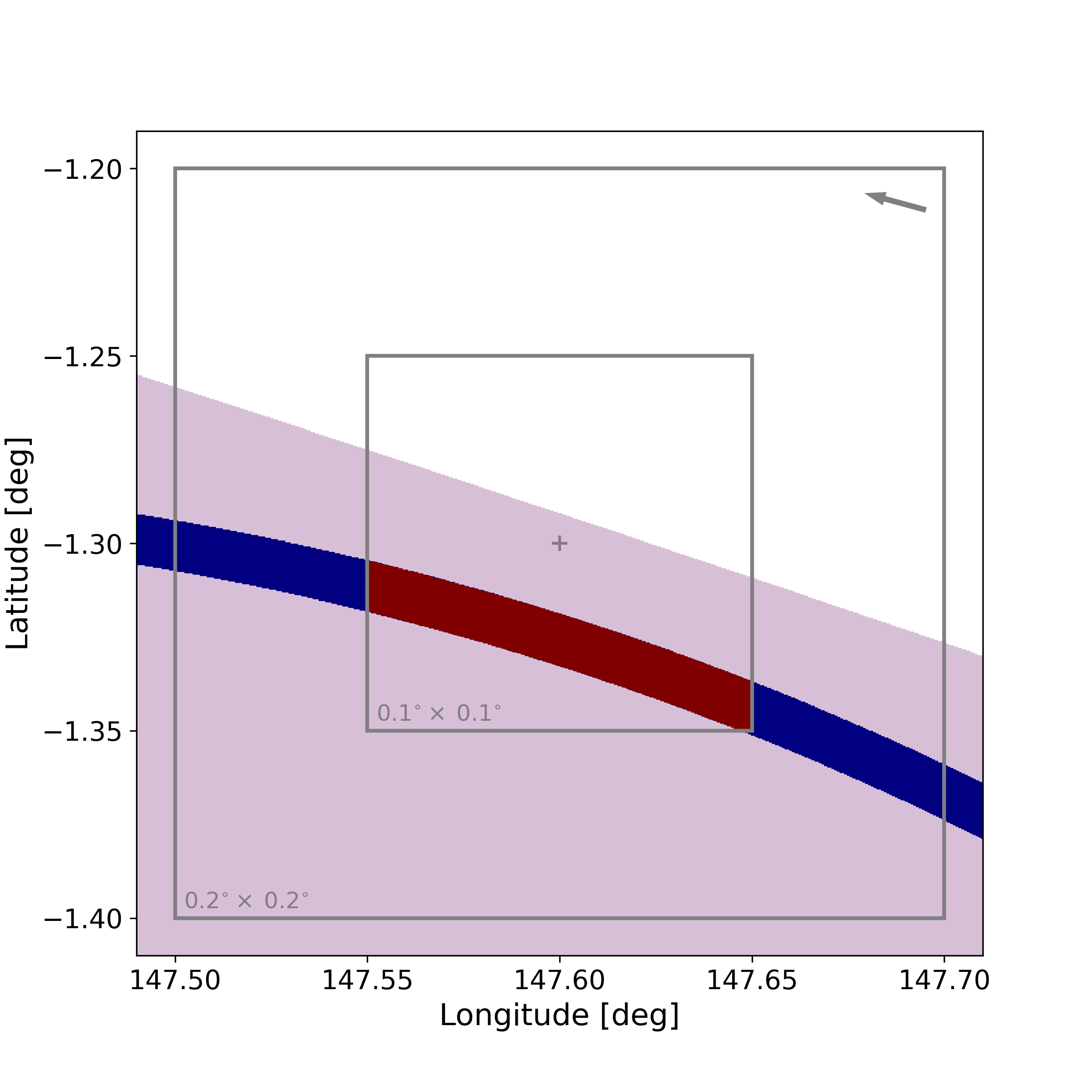}
\caption{The DoD-reported location is represented by the ``plus" at the center of the plot, with the $0.1^{\circ} \times 0.1^{\circ}$ box corresponding to the area allowed by the precision of the DoD-reported coordinates. AU MANU distance constraint (projected onto the surface of the ocean) is illustrated in blue, with the $\sim 16 \mathrm{\; km^2}$ portion that agrees with the DoD-reported coordinates (given their level of precision) highlighted in red. The purple region indicates the constraint given by the AU COEN seismic signal, which is fully consistent with the AU MANU distance constraint. The gray arrow indicates the direction that the meteor was traveling in, according to DoD data.} 
\label{fig:search_area}
\label{fig:map}
\end{figure}

We use a numerical procedure to check if the derived values for $r$ and $z$ correctly describe the shape of the second packet. The procedure imagines the blast from the fireball as a singular spherical wave emitted $z = 16.9 \pm 0.9 \mathrm{\; km}$ above a point ocean surface $r = 83.9 \pm 0.7 \mathrm{\; km}$ away from the AU MANU seismometer, traveling at $v = 315 \pm 2 \mathrm{\; m \; s^{-1}}$. The spherical wave intersects the ocean surface at times $t > (z/v_a)$ in circles with radii $r_{circ} = \arccos{[(z/(v_a t))]}$, carrying a fraction of the blast energy corresponding to the fraction of the spherical blast wave contacting the ocean surface, $f \propto (2 \pi r_{circ} \mathrm{d}r_{circ})/(4 \pi r^2)$. We assume an angle-independent reflection coefficient and the amplitude does not matter for the probability distribution in Figure \ref{fig:seismic}, given the arbitrary normalization. The procedure then considers each point on each ocean surface circle re-radiating, using the distance from the point to the AU MANU seismometer, $s$, to derive the additional travel time $(s / v_a)$ and the additional reduction in flux of $(4 \pi s^{-2})$. For each possible path, the travel time from the fireball to the point on the ocean surface, and from the point on the ocean surface to the AU MANU seismometer, are summed, and the relative flux delivered to the AU MANU seismometer is tabulated. This allows the relative fluxes for all paths with arrival times within each time bin to be summed. The result is the relative amplitude for the simulated signal, assuming perfect reflection and no transmission losses, as a function of time. The simulated second packet provides an excellent match to the rise, peak, and tail of the actual second packet, displayed in Figure \ref{fig:peak2}. This match confirms the blast location parameters derived using Equations \eqref{eq1} - \eqref{eq3}. 

We confirm the fireball location implied by AU MANU with the timing of the strong seismic signal, clearly distinct from background noise, detected by the AU COEN seismic station at 2014-01-08 18:23:53, presumably produced by the fireball. Given the sound speed, $v = 315 \pm 2 \mathrm{\; m \; s^{-1}}$, and the time difference between the AU COEN signal and the fireball, we derive a distance from AU COEN in the range, $1470 - 1490 \mathrm{\; km}$. This constraint is illustrated in purple in Figure \ref{fig:map}, and is fully consistent with the more narrow constraint derived from the AU MANU seismic signal. 


\section{Discussion}

The locations reported by the DoD in the CNEOS catalog are rounded to the tenths place in longitude and latitude. This suggests that the reported location, $(-1.3^{\circ}, 147.6^{\circ})$, corresponds to a square connecting $(-1.35^{\circ}, 147.65^{\circ})$, $(-1.35^{\circ}, 147.55^{\circ})$, $(-1.25^{\circ}, 147.55^{\circ})$, and $(-1.25^{\circ}, 147.55^{\circ})$. The area of this region is $\sim 120 \mathrm{\; km^2}$. The acoustic localization described here implies a distance of $83.9 \pm 0.7  \mathrm{\; km}$ from the AU MANU seismometer at $(-2.0432^{\circ},$ $147.366196^{\circ})$, and overlaps with the DoD-provided location only within a $\sim 16 \mathrm{\; km^2}$ region connecting $(-1.337^{\circ}, 147.650^{\circ})$, $(-1.350^{\circ}, 147.647^{\circ})$, $(-1.318^{\circ}, 147.55^{\circ})$, and $(-1.305^{\circ}, 147.55^{\circ})$. This reduces the uncertainty in the location of the fireball by a factor of $\sim 7.5$ in terms of area. This reduction in the geographic uncertainty of the IM1 fireball improves the search efficiency in the forthcoming ocean expedition to recover its fragments \citep{2022arXiv220800092S}. In addition, the direction in which the IM1 was traveling through the atmosphere fortuitously aligns with the distance constraint derived here, due to the serendipitous location of the AU MANU seismometer. This is beneficial for the ocean expedition search, as the fragments are expected to fall along the extrapolated post-fireball ground track trajectory of the meteor \citep{2022arXiv221200839T}. 

\section*{Acknowledgements}
We thank Tim Gallaudet and Steve Levine for providing crucial meteorological data used in this work. This work was supported in part by a grant from the Breakthrough Prize Foundation and by research funds from the Galileo Project at Harvard University. 





\bibliography{bib}{}
\bibliographystyle{aasjournal}



\end{document}